\documentclass[aps,prd,twocolumn,superscriptaddress,showpacs,floatfix]{revtex4}
\usepackage{graphicx}% Include figure files
\usepackage{dcolumn}% Align table columns on decimal point
\usepackage{bm}% bold math

\begin{document}

%\preprint{}

\title{Measurement of
Inclusive Spin Structure Functions of the Deuteron}
%with CLAS}

\newcommand*{\uvch}{University of Virginia, Department of Physics, Charlottesville, VA 22903, USA}
\newcommand*{\latech}{Center for Applied Physics Studies, Louisiana
Tech University, Ruston, LA 71272, USA}
\newcommand*{\cnuva}{Christopher Newport University, Newport News, VA 23606, USA}
\newcommand*{\jlab}{Thomas Jefferson National Accelerator Facility, 12000 Jefferson Avenue, Newport News, VA 23606, USA}
\newcommand*{\yerevan}{Yerevan Physics Institute, 375036 Yerevan, Armenia}
\newcommand*{\asuaz}{Arizona State University, Department of Physics and Astronomy, Tempe, AZ 85287, USA}
\newcommand*{\cmupa}{Carnegie Mellon University, Department of Physics, Pittsburgh, PA 15213, USA}
\newcommand*{\cuawdc}{Catholic University of America, Department of Physics, Washington D.C., 20064, USA}
\newcommand*{\cwm}{College of William and Mary, Department of Physics, Williamsburg, VA 23187, USA}
\newcommand*{\duke}{Duke University, Physics Bldg. TUNL, Durham, NC27706, USA}
\newcommand*{\edinburgh}{Department of Physics and Astronomy, Edinburgh University, Edinburgh EH9 3JZ, United Kingdom}
\newcommand*{\fiu}{Florida International University, Miami, FL 33199, USA}
\newcommand*{\fsu}{Florida State University, Department of Physics, Tallahassee, FL 32306, USA}
\newcommand*{\gwudc}{George Washington University, Department of Physics, Washington D. C., 20052 USA}
\newcommand*{\GLASGOW }{ University of Glasgow, Glasgow G12 8QQ, United Kingdom} 
\newcommand*{\frascati}{Istituto Nazionale di Fisica Nucleare, Laboratori Nazionali di Frascati, P.O. 13, 00044 Frascati, Italy}
\newcommand*{\genova}{Istituto Nazionale di Fisica Nucleare, Sezione di Genova
 e Dipartimento di Fisica dell'Universit\`a, 16146 Genova, Italy}
\newcommand*{\itep}{Institute of Theoretical and Experimental Physics, 25 B. Cheremushkinskaya, Moscow, 117259, Russia}
\newcommand*{\ipn}{Institut de Physique Nucleaire d'Orsay, IN2P3, BP 1, 91406 Orsay, France}
\newcommand*{\jmuva}{James Madison University, Department of Physics, Harrisonburg, VA 22807, USA}
\newcommand*{\knukorea}{Kyungpook National University, Department of Physics, Taegu 702-701, South Korea}
\newcommand*{\mmit}{Massachusetts Institute of Technology, Cambridge, Massachusetts  02139-4307}
\newcommand*{\nsuva}{Norfolk State University, Norfolk VA 23504, USA}
\newcommand*{\ohio}{Ohio University, Department of Physics, Athens, OH 45701, USA}
\newcommand*{\oduva}{Old Dominion University, Department of Physics, Norfolk VA 23529, USA}
\newcommand*{\rpi}{Rensselaer Polytechnic Institute, Department of Physics, Troy, NY 12181, USA}
\newcommand*{\rubltx}{Rice University, Bonner Lab, Box 1892, Houston, TX 77251, USA}
\newcommand*{\sphn}{CEA Saclay, DAPNIA-SPhN, F91191 Gif-sur-Yvette Cedex, France}
\newcommand*{\ucla}{University of California at Los Angeles, Department of Physics and 
Astronomy, Los Angeles, CA 90095-1547, USA}
\newcommand*{\connecticut}{University of Connecticut, Physics Department, Storrs, CT 06269, USA}
\newcommand*{\umma}{University of Massachusetts, Department of Physics, Amherst, MA 01003, USA}
\newcommand*{\umosk}{University of Moscow, Moscow, 119899 Russia}
\newcommand*{\unhdurham}{University of New Hampshire, Department of Physics, Durham, NH 03824, USA}
\newcommand*{\uppa}{University of Pittsburgh, Department of Physics and Astronomy, Pittsburgh, PA 15260, USA}
\newcommand*{\urva}{University of Richmond, Department of Physics, Richmond, VA 23173, USA}
\newcommand*{\usc}{University of South Carolina, Department of Physics, Columbia, SC 29208, USA}
\newcommand*{\utep}{University of Texas at El Paso, Department of Physics, El Paso, Texas 79968, USA}
\newcommand*{\vpsu}{Virginia Polytechnic and State University, Department of Physics, Blacksburg, VA 24061, USA}
\newcommand*{\ucny}{Union College, Department of Physics, Schenectady, NY 12308, USA}

\affiliation{\oduva}
\affiliation{\latech}
\affiliation{\genova}
\affiliation{\asuaz}
\affiliation{\sphn}
\affiliation{\ucla}
\affiliation{\cmupa}
\affiliation{\cuawdc}
\affiliation{\cnuva}
\affiliation{\cwm}
\affiliation{\connecticut}
\affiliation{\duke}
\affiliation{\edinburgh}
\affiliation{\fiu}
\affiliation{\fsu}
\affiliation{\gwudc}
\affiliation{\GLASGOW } 
\affiliation{\frascati}
\affiliation{\ipn}
\affiliation{\itep}
\affiliation{\jmuva}
\affiliation{\knukorea}
\affiliation{\mmit}
\affiliation{\umma}
\affiliation{\umosk}
\affiliation{\unhdurham}
\affiliation{\nsuva}
\affiliation{\ohio}
\affiliation{\uppa}
\affiliation{\rpi}
\affiliation{\rubltx}
\affiliation{\urva}
\affiliation{\usc}
\affiliation{\utep}
\affiliation{\jlab}
\affiliation{\ucny}
\affiliation{\uvch}
\affiliation{\vpsu}
\affiliation{\yerevan}


\author{J.~Yun}
\affiliation{\oduva}
\author{S.E.~Kuhn}
\email[Contact Author \ ]{skuhn@odu.edu}
\affiliation{\oduva}
\author{G.E.~Dodge}
\affiliation{\oduva}
%\thanks{}
\author{T.A.~Forest}
\affiliation{\oduva}
\affiliation{\latech}
\author{M.~Taiuti}
\affiliation{\genova}
\author{G.S.~Adams}\affiliation{\rpi}
\author{M.J.~Amaryan}\affiliation{\yerevan}
\author{E.~Anciant}\affiliation{\sphn}
\author{M.~Anghinolfi}\affiliation{\genova}
%\author{D.S.~Armstrong}\affiliation{\cwm}
\author{B.~Asavapibhop}\affiliation{\umma}
% ...

\collaboration{The CLAS Collaboration}
\noaffiliation

\date{\today}

\begin{abstract}
We report the results of a new measurement of spin structure functions
of the deuteron in the region of moderate momentum transfer
 ($Q^2$ = 0.27 -- 1.3 (GeV/c)$^2$) and final hadronic state mass 
in the nucleon resonance region ($W$ = 1.08 -- 2.0 GeV). 
%This experiment 
We scattered a 2.5 GeV polarized continuous
electron beam at Jefferson Lab off a dynamically polarized cryogenic solid state
target ($^{15}$ND$_3$) and detected the scattered electrons with the 
CEBAF Large Acceptance Spectrometer (CLAS). 
From our data, we extract the longitudinal double spin asymmetry
$A_{||}$ and the spin structure function $g_1^d$. 
Our data are generally
in reasonable agreement with existing data from SLAC where they overlap,  and they
represent a substantial improvement in statistical precision.
We compare our results with expectations for resonance asymmetries and
extrapolated deep inelastic scaling results. Finally, we evaluate the
first moment of the structure function $g_1^d$ and study its approach to
both the deep inelastic limit at large $Q^2$ and to the Gerasimov-Drell-Hearn
sum rule at the real photon limit ($Q^2 \rightarrow 0$). We find that the
first moment
varies rapidly in the $Q^2$ range of our experiment and
crosses zero at  $Q^2$ between 0.5 and 0.8 (GeV/c)$^2$, indicating the
importance of the $\Delta$ resonance at these momentum transfers.
\end{abstract}

%\keywords{}

\maketitle

%\section{}
\section{Introduction \label{intro}}

The nucleon spin structure functions
$g_1^{p,n}(x)$ and $ g_2^{p,n}(x)$
and their moments have been extensively studied
over the past two decades \cite{EMCfinal, SMCpLong, SMCd, E142Long, E143Long, E154g1,
E155Q2, E155xpaper, HERMESn, HERMESp}. 
At large momentum transfer ($Q^2 >> 1$ (GeV/c)$^2$) and final
state mass ($W > 2$ GeV), these data can be successfully described via perturbative QCD
(pQCD)
up to next--to--leading order (NLO) and give us access to the helicity-weighted distribution
functions $\Delta q (x)$ and $\Delta G (x)$ of quarks and gluons in the nucleon
\cite{StratmannNLO, GluckNLO, deFlorianNLO, E154NLO}. In this kinematic regime,
one can relate the first moments $\Gamma_1^N = \int_0^1 g_1^N(x)dx$ of the spin structure 
functions $g_1^N(x)$ ($N$ = $p$ or $n$)
to the fraction of the nucleon spin carried by the quark helicities and, via
the famous Bjorken sum rule \cite{BjSR,BjSR2}, to the weak axial form factor $g_A$.
%Existing data show that only 20\% -- 30\% of the nucleon
%spin is carried by the quark helicities. The data deviate by nearly 5 standard deviations from the
%individual predictions for $\Gamma_1^{p,n}$ for the proton and the
%neutron by Ellis and Jaffe \cite{EJSR}, 
%which were based on the assumption of approximate SU(6)
%symmetry and negligible contributions from strange quarks. However, the data
% confirm the Bjorken sum rule for the 
%proton--neutron difference, $\Gamma_1^p - \Gamma_1^n = g_A/6$ (plus
%pQCD corrections).

At lower momentum transfers, $Q^2 \approx 1$ (GeV/c)$^2$,
corrections proportional to powers of
$1/Q^2$ develop due to higher twist and target mass effects 
\cite{SSFtwist4,SSFtwist3, BalitskySSF}
in addition to the logarithmic
$Q^2$ dependence predicted by pQCD. 
%Through the method of
%operator product expansion (OPE) \cite{OPE1, OPE2, SSFope},
%one can relate these higher twist effects to quark-gluon correlations in the nucleon
%wave function via higher moments of the structure functions $g_1$ and $g_2$.
As $Q^2$ decreases, an increasing part of the kinematic range $x = 0$  --  $1$ 
lies in the region
of resonant final states ($W<2$ GeV), which begin to dominate the spin structure
functions. They become less positive
 (or more negative in the case of the neutron),
%eventually change sign, 
in particular
in the region of the $\Delta$ resonance. Data in this region on 
structure functions and on the (virtual) photon
asymmetries $A_1$ and $A_2$ for the proton and the neutron,
\begin{eqnarray}
A_1 = \frac{\sigma_{1/2}-\sigma_{3/2}}{\sigma_{1/2}+\sigma_{3/2}}
=  \frac{g_1 -  g_2/\tau }{F_1},
\nonumber \\
A_2= \frac{\sigma_{LT}}{\sigma_{1/2}+\sigma_{3/2}} 
=  \frac{g_1 + g_2}{\sqrt{\tau} F_1}, 
\label{A1eq}
\end{eqnarray}
can help us unravel the spin-isospin
structure of resonance transition amplitudes and their interference with 
each other and with non-resonant
terms. 
We can also test whether the observed duality between
unpolarized deep inelastic and resonant structure
 functions \cite{BloomGilman,NiculescuDual1, NiculescuDual2}
 is realized for spin structure functions as well \cite{CarlsonDualg1,SimulaDualg1}.
Here, $\sigma_{1/2}$ and $\sigma_{3/2}$  are the
(virtual)  photon absorption cross sections for total
(photon plus nucleon) helicity 1/2 and 3/2 and $\sigma_{LT}$ is the
 longitudinal--transverse interference cross section,
$F_1$ is the unpolarized structure function,
and $\tau = \nu^2/Q^2$ with $\nu = E - E'$ being the energy
loss of the scattered electron.

Due to the dominance of the resonances at low $Q^2$, 
the integrals $\Gamma^{p}$ and 
$\Gamma^{d} \approx (\Gamma^{p} + \Gamma^{n})/2$
(which are positive in the scaling region of high $Q^2$) decrease 
rapidly and become negative as $Q^{2}$ approaches zero. 
In the limit $Q^2 \rightarrow 0$, the first moments 
for the proton and the neutron are
constrained by the Gerasimov-Drell-Hearn Sum Rule \cite{Gerasimov,DrellHearn},
which predicts that
\begin{equation}
\Gamma_{1}^{N}(Q^2) \rightarrow
{{Q^{2}} \over {16 \pi^{2} \alpha}}
\int_{\nu_{thr}}^{\infty} \left(\sigma_{1/2}-\sigma_{3/2} \right) \frac {d\nu} {\nu}
=-{{Q^{2}} \over {8M^{2}}} \kappa_{N}^{2} .
\label{GDHeq}
\end{equation}
Here, $\alpha$ is the fine structure constant and $M$ and $\kappa_N$ are the
mass and anomalous magnetic moment of the nucleon, respectively. 
Since the GDH sum rule is negative, the integrals $\Gamma_{1}^{p,d}(Q^{2})$
must have a negative slope at $Q^{2} = 0$ and then
change rapidly at low $Q^{2}$ to meet the positive experimental results in the DIS region.

%We can extract higher twist matrix elements from moments of spin structure
%functions and their dependence on $Q^2$ \cite{JiTwist4}.
%Near the photon point, we can test predictions by chiral perturbation theory ($\chi$PT) 
%\cite {MeissnerChPT,JiChPT} for the first
%and second derivative of the integral with respect to $Q^2$.
%Ultimately, we can study the evolution of the generalized Gerasimov-Drell-Hearn Sum Rule
%\cite{JiGDH} over the whole range of momentum scales, from
%$Q^2 = 0$ where hadronic
%degrees of freedom (nucleons, pions and resonances)
%are relevant, up to the deep inelastic region which is dominated by
%asymptotically free quarks. 

So far, only phenomenological models for $\Gamma_{1}(Q^{2})$ covering the
whole range of $Q^2$
 exist \cite{Anselmino,SofferGam1p,SofferGam1n,BurkertLi,BurkertIoffe,BurkertIoffe2,
JiGam1Q2}.
These models are constrained to reach the large--$Q^2$ asymptotic value of the integral
as measured by deep inelastic data and to approach zero at
the photon point with a slope given by the Gerasimov-Drell-Hearn sum rule,
Eq.~(\ref{GDHeq}). 
The authors of
Refs. \cite{SofferGam1p,SofferGam1n} use a simple parametrization of the integral
$\Gamma_{1+2} = \int [g_1(x)+g_2(x)] dx$ to interpolate between these two points, and
then subtract the integral over $g_2$ which is given by the Burkardt-Cottingham sum rule
\cite{BCSR}. The approach taken in Refs. \cite{BurkertLi,BurkertIoffe,BurkertIoffe2} uses a parametrization
of existing resonance data 
and a vector meson dominance (VMD) inspired
interpolation of the remaining integral strength at the two endpoints.

For a complete picture of spin structure functions and their moments,
one needs information on both the proton
and the neutron. 
Since free neutron targets are impractical, deuterium (as in the experiment
described here)
 or $^3$He targets
are used instead. An unambiguous extraction of neutron spin structure
functions from nuclear ones is less straightforward 
in the resonance region than in the deep inelastic
regime; however, the integrals $\Gamma_1^N$ are much less affected by uncertainties from
Fermi motion, off-shell effects, and other nuclear corrections \cite{Umnikov,MelThom,Ciofi}.
In particular, studies \cite{Ciofi2,Arenhovel} show that the integral $\Gamma_1^d$
for the deuteron from pion threshold on up is very close to the incoherent sum of 
proton and neutron integrals, once a correction for the deuteron D-state has been applied.

So far, only very limited 
spin structure function data exist in the region of low to moderate $Q^2$ and $W$ 
\cite{E130Res,E143Res}, especially on the deuteron. 
%We report the first results on the deuteron from a 
A large program is underway at Jefferson Lab to map out the
entire kinematic region $Q^2 \approx 0.05$  -- $ 5$ (GeV/c)$^2$ and $W \leq 3$ GeV.
This program consists of measurements on $^3$He (in Hall A) and on proton and
deuteron targets with CLAS (the EG1 collaboration in Hall B). First results from
CLAS \cite{EG1apiplus} and Hall A \cite{HallAGDH1} have already been published. 

In the present paper, we present results on the deuteron from the first EG1 run in 1998, 
in which we measured
double spin asymmetries $A_{||} = D(A_1 + \eta A_2)$
on deuterium with a beam energy of 2.5 GeV.  ($D$ and $\eta$ are
kinematical factors, see Sec.~\ref{res}.) These data cover a range in
$Q^2$ from 0.27 -- 1.3 (GeV/c)$^2$ and final state mass 
in the resonance region ($W$ = 1.08 -- 2.0 GeV).
The remaining data set
from EG1 is presently under analysis and will increase both the kinematic coverage and
the statistical precision of our data significantly.

In the following,
we give some details on the experiment (Sec.~\ref{exp}) and its analysis
(Sec.~\ref{analy}). We present our results on the 
deuteron spin asymmetry $\left(A_1^d + \eta A_2^d \right) (W,Q^2)$,
the structure function $g_1^d(x,Q^2)$ and its first moment $\Gamma_1^d(Q^2)$ (Sec.~\ref{res}),
and
conclude with a summary and outlook (Sec.~\ref{sum}). 
\section{Experimental Details \label{exp}}

The data described in this
paper were collected during a three-month run in 1998, as part of the
EG1 run group in Jefferson Lab's Hall B.  A polarized electron beam with  
2.5 GeV beam energy
was scattered off a deuterated ammonia ($^{15}$ND$_3$) target
that was dynamically polarized along the beam direction.
The average beam current of 2.5 nA corresponded to an instantaneous luminosity of 
$0.4 \times 10^{34}$ cm$^{-2}$s$^{-1}$. The beam polarization was measured 
periodically with a M\o ller polarimeter and the average beam polarization was
72\%.

We used the CEBAF Large Acceptance Spectrometer (CLAS) 
to detect the scattered electrons.
The CLAS detector \cite{CLASoverview} is built around six superconducting
coils that produce a toroidal magnetic field.  The orientation of the
magnetic field can be chosen so that  electrons are bent either toward
(inbending) or away from the beam line (outbending).
The target was placed 55 cm upstream from its
normal location in the center of CLAS to lower the angular
threshold for electron detection and thus decrease the lower limit on the
momentum transfer.  Inbending
electrons were detected down to a minimum polar angle of
14$^\circ$.  During this experiment
the geometry of the target excluded particle tracks with a polar angle between 50$^\circ$
and 75$^\circ$.  The phi--acceptance is approximately 85\%, limited mainly by
the torus coils.  

The CLAS detector package consists of three layers of
drift chambers for track reconstruction,
one layer of scintillators for time--of--flight measurements, forward Cerenkov
counters for electron--pion discrimination, and electromagnetic
calorimeters to identify electrons and neutral particles.
A coincidence between the Cerenkov and the
calorimeter triggers the data acquisition.  Electron particle identification
is accomplished using the Cerenkov detector and the distribution of energy
deposited in the calorimeter.
 The large acceptance of CLAS ($\approx$ 1.5 sr for electrons)
and its large kinematic coverage
offset the limited luminosity that can typically be reached with
polarized solid state targets (of order $10^{35}$ cm$^{-2}$s$^{-1}$ at best),
and allowed us to collect data for the entire $W$ and $Q^2$
range simultaneously.

The longitudinally
polarized target was designed to fit within the 1 m central bore
of the CLAS~\cite{target}. A pair of superconducting Helmholtz coils
provided a 5 T magnetic field along the direction of the electron beam.
The magnetic field was uniform to better than $1 \times 10^{-4}$ in the
center of the target over a length of 2 cm and a diameter of 2 cm. The
ammonia crystals were contained within a plastic cylindrical cell 1 cm
in length and 1.5 cm in diameter.  The cell was immersed in a liquid He bath
maintained at approximately 1 K by a $^4$He evaporation refrigerator.  The
cell was mounted on a target insert that also held a NH$_3$ cell, as well
as a $^{12}$C and an empty cell. The latter cells were used to study the 
dilution of the measured asymmetries by events from unpolarized target
constituents (see Sec.~\ref{dilution}). 
The deuterons in the target were polarized using the
Dynamic Nuclear Polarization (DNP) technique \cite{Abragam, Crabb} with
140 GHz microwaves.  The polarization of the target was monitored on-line
using the NMR technique. The NMR results were not used for our final
analysis; instead, we extracted the product of
beam and target polarization directly from our data, as described in Sec.~\ref{dilution}.
%However, the NMR coils were placed on the outside of the deuteron cell and
%may have been sensitive to material with a higher average polarization than
%that exposed to the electron beam.  For that reason we used the product of
%beam and target polarization as determined in Section \ref{dilution}
%for the data analysis.
The beam was rastered on the ND$_3$ target, although not over the full face
of the target.
The deuteron polarization suffered from this incomplete raster and from
inadequate microwave power and
ranged
from approximately 10\% to 25\%.
%new
All data were taken with the target polarization along the beam direction,
without reversal of the target polarization. The beam helicity was reversed
every second.
%endnew

\begin{figure}[htbp]
\includegraphics[width=8cm]{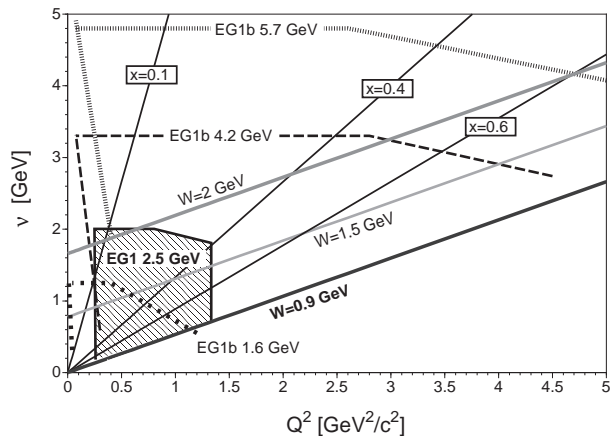}
\caption{
Kinematic coverage of the data described in this paper (``EG1 2.5 GeV'')
together with the kinematic range of the second run of EG1
(``EG1b'' at 1.6 GeV, 4.2 GeV, and 5.7 GeV). The heavy solid lines indicate
the elastic peak (``$W$ = 0.9 GeV''), the location of the S$_{11}$ resonance
(``$W$ = 1.5 GeV'') and the deep inelastic limit (``$W$ = 2 GeV''). Also shown
are the kinematic lines for three representative values of $x$.
\label{kine}}
\end{figure}

During the 1998 run,
we collected 300 million triggers for an integrated beam charge of about 0.4 mC.
From this sample,
100 million electron events passed the cuts described in Sec.~\ref{select}. These
events covered
a kinematic region from the quasielastic region ($W \approx 0.94$ GeV)
to the edge of
the deep inelastic region ($W = 2$ GeV) and for $Q^2$ = 0.27 -- 1.3 (GeV/c)$^2$.
This kinematic coverage is shown in Fig.~\ref{kine}, together
with the coverage of the second part of the EG1 experiment.

\section{Data analysis \label{analy}}

The goal of our analysis was to first determine the double spin asymmetry
\begin{equation}
A_{||} = \frac{\sigma ^{\uparrow \downarrow} - \sigma ^{\uparrow \uparrow}}
{\sigma ^{\uparrow \downarrow} + \sigma ^{\uparrow \uparrow}}
\label{Aparadef}
\end{equation}
for each kinematic bin and then to extract the physical quantities of interest, the
virtual photon asymmetries $A_1^d + \eta A_2^d$ and 
the structure function $g_1^d$,
from the results. Here, $\sigma ^{\uparrow \downarrow} $ stands for the
differential electron scattering cross section with the target and
electron spin pointing in opposite directions along the beam and
correspondingly $\sigma ^{\uparrow \uparrow}$ for parallel target
and electron spin.

\subsection{Data selection \label{select}}

For the present analysis, we selected data runs taken with a torus current of +2250 A
(inbending electrons) and target
polarization parallel to the beam direction. The data were taken with two slightly different
beam energies, 2.494 GeV and 2.565 GeV, due to a change of the
accelerator configuration. We separated our sample into
four different ``run groups'', two each with beam energy 2.494 GeV and 2.565 GeV. 
%new
Each run group corresponds to a contiguous set of runs with the same target
material and approximately constant target parameters and running conditions.
%endnew 
Only
runs with stable beam and detector performance were included in our sample. The 2.565 GeV
groups also contained carbon target runs 
that were used to determine the dilution factor (see Sec.~\ref{dilution}). 
We analyzed events with
scattering angles from about 14$^{\circ}$ to 50$^{\circ}$ and scattered electron energies
from 0.5 GeV to 2.5 GeV.

The data were sorted according to the helicity of the electron beam. During our run,
the beam helicity followed a ``pseudo--random'' pattern of helicity pairs, where the
first ``bucket'' (of 1 second length) of each pair was given random helicity and the second
its complement. We matched the sequence of helicity bits for each event with the
pattern sequence recorded in helicity scalers and discarded pairs for which the
helicity assignment was inconsistent. We also discarded pairs with significantly
different (by more than 10\%)
beam intensity in the two buckets (due to beam fluctuations or trips). The final
data sample contained only matched pairs of buckets with stable running
conditions.

All events were accumulated in small bins of $W$ ($\Delta W = 0.02$ GeV)
and $Q^2$ ($\Delta Q^2 / Q^2 \approx 20\%$), separately for both beam
helicities.
(The data on asymmetries and $g_1$, shown in Sec.~\ref{res}, are weighted
averages of several such bins).
 In addition, we also accumulated the integrated beam charge for each
of the helicity buckets (corrected for deadtime)
to normalize the helicity-sorted counts in each bin.  We found
that on average, there was a 0.3\% difference between the integrated charge for the
two opposite helicities, possibly stemming from the sensitivity of the photocathode
in the polarized source
to small remaining linear polarization components 
or beam motions of the photo--ionization laser beam. Our
normalization method removed the effect of this asymmetry, and it was further suppressed
by reversing the relative sign between the helicity at the cathode and in the experimental
Hall (through spin--precession in the injector and the accelerator).

\subsection{Electron cuts}

We selected electron events by first requiring a negative track with matching
signals in the time--of--flight (ToF) scintillators, the Cerenkov counters (CC),
and the electromagnetic calorimeter (EC). In the presence of several 
such tracks, the track with the shortest flight time was selected as the
electron candidate.
Some additional cuts on the track vertex along the beam line removed
events from the entrance and exit windows of the polarized target
chamber, as well as badly reconstructed tracks.

\begin{figure}[htbp]
\includegraphics[width=8cm]{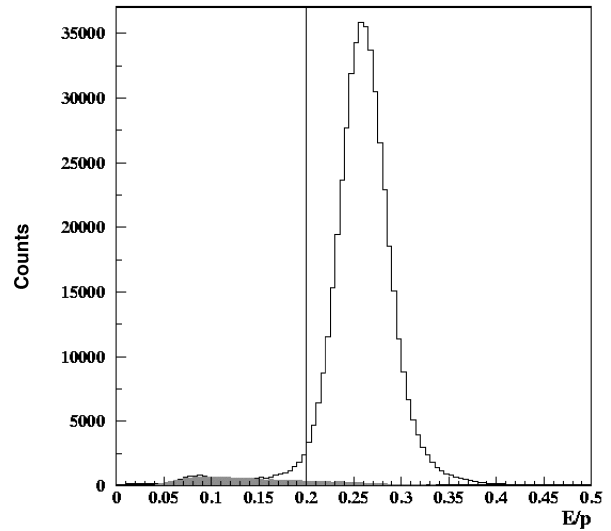}
%\special{psfile=Eoverp2.ps hscale=80 vscale=70 
%hoffset=0        voffset=-500
%}
\caption{
Spectra of the ratio of measured 
energy, $E$ (in GeV),  in the electromagnetic
calorimeter over the 
track momentum, $p$ (in GeV/c),
for electrons (open histogram)
and pions (shaded area). The vertical scale is arbitrary. Both
spectra have been cross--normalized at low $E/p$. Events
above the indicated threshold are identified as electrons.
\label{eoverp}}
\end{figure}

We used information from the CC and the EC to further
separate electrons from negative pions. We required a signal 
in the CC that exceeded 50\% of the average signal
 for a single photo--electron.
%corresponding
%to more than 0.5 photo--electrons in the CC. 
Furthermore, we required that
the energy measured in the EC exceeded
20\% of the candidate electron momentum (the average sampling fraction
of the EC was 27\%). A typical example for the ratio of sampled EC energy
over momentum is shown in Fig.~\ref{eoverp}. The open histogram shows
events that passed all other electron cuts (including the CC cut).

We also collected a sample of $\pi ^-$ events with no signal above
threshold in the CC. As shown by the shaded area in Fig.~\ref{eoverp},
the $E/p$ spectrum associated with $\pi ^-$ events is strikingly
different from the electron spectrum.
Under the conservative assumption that {\em all} events below a
$E/p$ ratio of 0.15 came from pions, we cross--normalized the two
spectra below that point and estimated the remaining pion contamination of our
electron sample by the ratio of the two integrated spectra above our
cut of $E/p > 0.2$. For all kinematics studied, this remaining contamination
turned out to be less than 1\%.

The reconstructed momenta of the scattered electrons were corrected 
for effects from
unknown torus field distortions and slight drift chamber misalignments.
We used NH$_3$ runs taken interleaved with the ND$_3$ ones to
determine the correction factor
by optimizing the position and width of the elastic peak ($W=0.938$ GeV)
for all scattering angles $\theta$ and $\phi$. The resulting corrections
were of the order 0.1\% on average.

\subsection{Dilution and polarization \label{dilution}}

The double spin asymmetry $A_{||}$ can be extracted from the 
count rate asymmetry (normalized by the integrated beam charge) after
accounting for the dilution from unpolarized target constituents and 
the beam ($P_b$) and target ($P_t$) polarization:
\begin{equation}
A_{||}^{meas} =\frac{1}{DF \cdot P_b \cdot P_t} \frac{N^+/Q^+ - N^-/Q^-}{N^+/Q^+ + N^-/Q^-}, 
\label{apararaw}
\end{equation}
where $N^{+,\thinspace -}$ are the counts and $Q^{+,\thinspace -}$ 
are the integrated
beam charge for positive and negative helicity.

We determined the dilution factor $DF$ in Eq.~(\ref{apararaw}) by approximating the
contribution to the count rates from unpolarized target constituents (target foils,
LHe coolant and $^{15}$N in ammonia) with the spectra taken on the carbon target.
Some components of these two targets were the same ({\it e.g.}, the LHe coolant and
foils were present for the carbon target as well), and carbon, nitrogen, and even
$^4$He have similar
binding energies per nucleon and Fermi momenta, suggesting that their inclusive electron
scattering spectra are similar after correcting for the total number of target
nucleons. (This assumption has since been verified to better than
3\% with dedicated runs on a pure $^{15}$N target during the second part of EG1).

To account for the different number of nucleons in each target and different overall
target thicknesses, we cross--normalized the carbon  target spectra to the ammonia
target spectra. We determined a normalization constant $A$ such that the two spectra
had the same number of counts below a cut--off missing mass $W_{cut}$, well below
the quasi--elastic peak.
The cut--off ranged from $W_{cut} = 0.835$ GeV at $Q^2 = 0.3$ (GeV/c)$^2$ to
 $W_{cut} = 0.5$ GeV at $Q^2 = 1.2$ (GeV/c)$^2$, and was chosen so that
the deuteron contribution was negligible, according to a Monte Carlo simulation
of the deuteron wave function.

\begin{figure}[htbp]
\includegraphics[width=7cm]{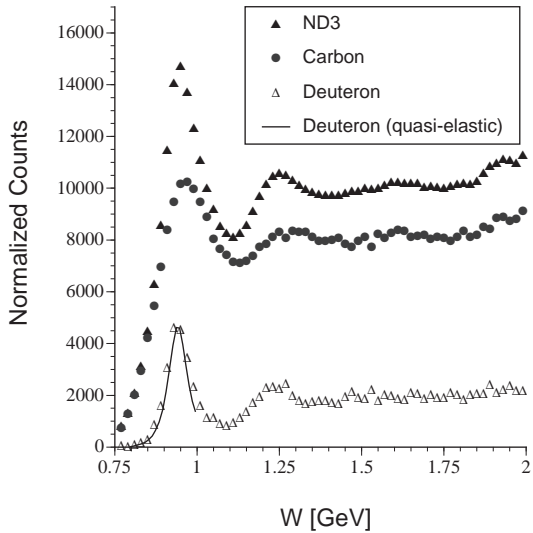}
\caption{
Spectra of counts {\it vs.} final state mass $W$ from the polarized ND$_3$ target (solid
triangles)
and carbon target (circles) runs for the range $Q^2 = 0.5 \pm 0.1$ (GeV/c)$^2$.
The spectra have been cross-normalized at low $W$.
The deuteron spectrum (open triangles) is the difference between these two spectra. 
In the quasi-elastic peak region, it agrees
well with a simulation  using the Paris wave function
for the deuteron (solid line).
\label{dilu}}
\end{figure}

The dilution
factor can then be written as
\begin{equation}
DF = \frac{N_{ND_3} - A \cdot N_{C}}{N_{ND_3}} ,
\label{dilfac}
\end{equation}
where the numerator is the count rate due to deuterium alone. The results of this
method for an intermediate $Q^2$ bin are shown in Fig.~\ref{dilu}. The normalized
carbon spectrum (circles) has been subtracted from the ammonia spectrum (solid
triangles)
to yield the deuteron spectrum (open triangles). The line indicates the result of our
Monte Carlo
simulation of the deuteron spectrum alone, which is based on quasi-elastic
scattering (PWIA) and the Paris wave function~\cite{Paris} for the deuteron. The dilution
factor for our experiment was around $DF \approx 0.2$.

The second ingredient needed in Eq.~(\ref{apararaw}) is the product of beam
and target polarization.
We measured both the beam polarization (with a M\o ller polarimeter) and
the target polarization (using NMR) individually during the run.
However, due to the small amount of target material and its inhomogeneous
exposure to the electron beam, the NMR results were not very precise and
reliable.
Instead, we determined directly the
product $P_b \cdot P_t$ by extracting it from the measured asymmetry in the
quasi--elastic region. For this purpose, we used 
%both fully reconstructed quasi--free
%proton scattering events $d(e,e'p)$ and 
inclusive quasi--elastic events $d(e,e')$ in the
range 0.85 GeV$\leq W \leq$ 1.0 GeV.

The asymmetry $A_{||}$ for elastic scattering from protons and 
neutrons can be calculated from known nucleon form factors with very little systematic
uncertainty (less than 1-2\% in our kinematic region). 
%For the inclusive 
%$d(e,e')$ events, 
We used our simulation of the deuteron wave function to calculate
the expected asymmetry for inclusive quasi--elastic scattering
within our kinematic cuts, which differed only slightly from the 
cross section--weighted average
of the proton and neutron asymmetries. We used the dilution factor determined by the
method described above to extract the product $P_b \cdot P_t$. 
%For the
%exclusive $d(e,e'p)$ events, we used tight cuts on the electron and proton
%angles and momenta, leading to an enhancement of the deuteron
%contribution. The dilution from quasi--elastic scattering on heavier
%target nuclei
%factor 
%in this case was determined from the
%``side wings'' of the distribution of $\phi_e - \phi_p$ which had a narrow
%deuteron peak at 180$^\circ$ on top of a broader background from other
%target nuclei.

\begin{figure}[htbp]
\includegraphics[width=8cm]{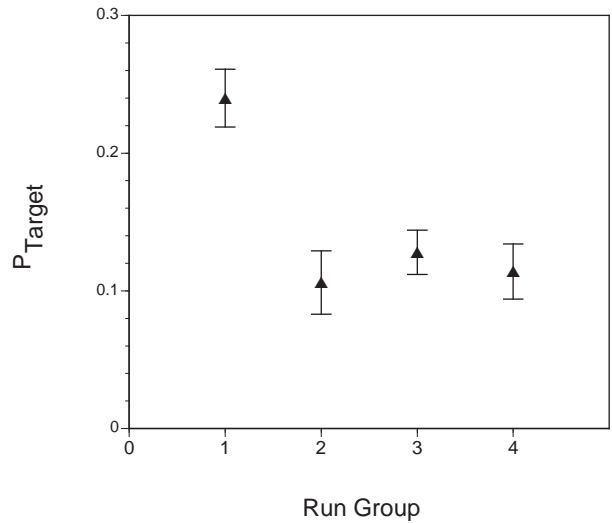}
\caption{
Average target polarization for each of the four run groups determined
by dividing the values of the product $P_b \cdot P_t$ extracted from the
quasi--elastic asymmetry by the beam polarization measured with
a M\o ller polarimeter. The  target polarization decreases
over time due to
beam exposure.  
%The online results from
%the NMR measurements (crosses) are compared with the results from the 
%inclusive analysis of the quasi-elastic peak (circles) and the exclusive
%d(e,e'p) analysis (triangles). Due to the uneven radiation exposure of the
%target material, the true polarization falls more rapidly with time than that
%measured by the NMR method.
\label{pol}}
\end{figure}

%new
Due to the large kinematic coverage of CLAS, data on the quasi--elastic asymmetries
were collected continuously and simultaneously with the inelastic asymmetry data.
The extracted average polarization product $P_b \cdot P_t$ for each of the four run
groups is therefore a faithful representation of the running conditions for that group,
with minimal systematic uncertainties. 
Our results 
are shown in Fig.~\ref{pol}, where we divided the product
$P_b \cdot P_t$ by the measured beam polarization to extract the
target polarization. 
The results for each of the individual run groups
have statistical errors on the order of 13\%, which were included in the total statistical
error of the asymmetries from each run group. The final results for the inelastic
asymmetries are statistically weighted
averages from the four run groups, with a contribution to their
statistical errors from the polarization product of about 6.7\% 
of their values.
%endnew

\subsection{Other backgrounds}

After dividing out the dilution factor and 
the beam and target polarizations in Eq.~(\ref{apararaw}),
we corrected the extracted asymmetry for  additional background contributions. These
include contamination of the scattered electron sample by negative pions and pair-produced
electrons, as well as contributions from polarized target constituents other than deuterium.

We already discussed the contribution from pions misidentified as electrons, which was
less than 1\% in all cases. A more important contribution comes from electrons that are
decay products of neutral pions (either through the Dalitz decay
$\pi^0 \rightarrow \gamma e^+ e^-$ or pair conversion of decay photons). The rate of
electrons from these decays was estimated using the Wiser fit~\cite{Wiser} 
for pion photoproduction
and tested against the Monte Carlo code ``PYTHIA''. 
%new
We also measured directly the rate of positron production in each kinematic bin
(again making use of the large acceptance of CLAS for both positively and negatively
charged particles).
This rate should be equal to that
of electrons from charge-symmetric decays and was found to agree well with
the Wiser fit. The asymmetry for positrons was found to be consistent with
zero and in any case no larger than the asymmetry for electron scattering events.
%endnew 
We used a parametrization of
our results to estimate the fraction of detected electrons coming from these decays.
This fraction was typically 1\% for most of the kinematic region, but increased
up to 20\% at the highest $W$ values. 
%The asymmetry for these pair--symmetric
%events was found to be consistent with zero. 
We corrected our data for
this background by
applying a further dilution factor to our asymmetries.
Since we could not exclude a small non--zero asymmetry for these events,
we assumed a systematic uncertainty equal to the size of this correction.

The nitrogen in our dynamically polarized ammonia target carries a small residual 
polarization, which leads to a partially polarized bound proton in $^{15}$N. 
Possible additional
polarized target species include isotopic impurities of $^{14}$N and $^1$H.
Extensive experience with similar targets at SLAC~\cite{E143Long} shows
that the corresponding corrections to the asymmetry are at most a few percent.
We included the uncertainty due to these contributions in our systematic error.

Another potential contribution to the measured asymmetry comes from parity-violating
electron scattering off all target constituents. However, at the low momentum
transfers of our experiment, the expected asymmetry is less than $10^{-4}$
\cite{Prescott} and can
be treated as another (small) systematic uncertainty.

\subsection{Radiative corrections and models}

The final step in the extraction of the desired ``Born asymmetry''
%\begin{equation}
$A_{||}$ 
% = \frac{\sigma ^{\uparrow \downarrow} - \sigma ^{\uparrow \uparrow}}
%{\sigma ^{\uparrow \downarrow} + \sigma ^{\uparrow \uparrow}}
%\label{Aborn}
%\end{equation}
requires correcting the measured asymmetry for higher--order electromagnetic
processes (internal radiative corrections)
 and electron energy loss through bremsstrahlung in the target before or after
the scattering (external radiative corrections).
These radiative corrections were applied separately to the numerator and the
denominator of Eq.~(\ref{Aparadef}), which yields an additive ($A_{RC}$) and a multiplicative
($F_{RC}$) correction term:
\begin{equation}
A_{||} = A_{||}^{meas}/F_{RC} + A_{RC} .
\label{ARCFRC}
\end{equation}
Here, the factor $1/F_{RC}$ represents the increase of the
denominator in Eq.~(\ref{Aparadef}) due to the radiative elastic and quasi-elastic tails
that act like an additional dilution of the inelastic events. Correspondingly,
the statistical error of the final result was scaled up by $1/F_{RC}$ as well.

Both components ($F_{RC}$, $A_{RC}$) were determined by running the
code ``RCSLACPOL'' developed at SLAC~\cite{E143Long}. This code
uses parametrizations of all relevant input quantities (structure functions and
form factors), as well as a model of our target, to calculate both fully radiated and
Born  cross sections and asymmetries. It is based on the approach developed by 
Kuchto and Shumeiko~\cite{Kukhto}
for the internal corrections and by Tsai~\cite{Tsai}
for the external corrections, including
the radiative depolarization of the beam due to external bremsstrahlung.

We used parametrizations of the world data on polarized and unpolarized
structure functions and elastic form factors as input for the radiative
correction code and to extract physics quantities of interest from the
measured asymmetries. These parametrizations
are described in~\cite{E155Q2} and
are based on fits to unpolarized structure function data from
NMC~\cite{NMCF2} and SLAC~\cite{BodekSF, E140, NE11delta, E143R}
and polarized structure function data from
SLAC~\cite{E130Res, E142Long, E143Long, E154g1, E155Q2, E155xpaper},
CERN~\cite{EMCfinal,SMCpLong, SMCd}, and 
HERMES~\cite{HERMESn, HERMESp}.  The nucleon form factors
were taken from Ref.~\cite{Bosted} with updated values for the ratio
$G_{Ep}/G_{Mp}$ from the recent Jefferson Lab experiment~\cite{Jones}.
For the asymmetries $A_1$ and $A_2$ in the resonance region, we used
parametrizations of resonance transition amplitudes from Ref.~\cite{BurkertLi}
(in the form of a computer code named ``AO'')
and Ref.~\cite{dre99} (``MAID'') together with
a fit of the SLAC data~\cite{E143Res}. We also included our own
preliminary asymmetry data in these fits. 
All fits were varied within reasonable errors or replaced with alternative existing
fits to study the systematic dependence of our final results on these
parametrizations.

\subsection{Systematic errors}

The total systematic error on our data ranges from 25\% to 50\% of
the statistical error for the asymmetries and from 35\% to 50\% of
the statistical error for the structure function $g_1^d$.
The leading contributions to these systematic uncertainties come from
radiative corrections (40--50\% of the total systematic error on average),
uncertainties in the unpolarized structure functions
needed to extract final physics results (also 40--50\% of the total),
and the dilution factor (about 40\%).
We also considered the effect of finite resolution and errors
in the measured kinematic variables (about 10\% of the total). 
At higher $Q^2$ and especially
higher $W$, pair-symmetric decay electrons also contributed significantly
to the overall systematic uncertainty (15--20\% averaged over all kinematic bins
and most of the systematic error at the kinematic limit).
Finally, for the extraction of the spin 
structure function $g_1^d$ and its integrals, some model assumption about
the virtual photon asymmetry $A_2$ is needed 
(see Sec.~\ref{res}) and leads to a further
systematic error (up to 50\%).

We accounted for each of these systematic errors by changing a relevant input
parameter or model, and then repeating the entire analysis up to the final results,
including the integrals of $g_1^d$ over the measured region.
We took the error as the deviation of the alternative results from the standard
analysis. We added all uncorrelated systematic errors in quadrature. The final
systematic errors are shown in the data tables in Sec.~\ref{res}.

For the radiative correction errors, we varied all input models and parametrizations
for the radiative code, including polarized and unpolarized structure functions,
form factors, and the target model, within realistic limits. We also checked the
accuracy of the peaking approximation by comparing the results with those
from a full integration without approximations.

Similarly, we varied the models for the unpolarized structure functions 
$F_1^d$ and $R ={ \sigma_L \over \sigma_T}$, which entered the extraction
of $g_1^d$ and the asymmetry $A_1^d + \eta A_2^d$ from our data (see Sec.~\ref{res}).
We used different fits of the world 
data~\cite{BodekSF,E143R,RRicco}, and studied their effect
on the final physics results. In the case of the polarized structure function
$g_1^d$ and its integrals, we also varied the model for the asymmetry $A_2^d$
from $A_2^d = 0$ to the prediction by the ``MAID'' code and a simple
parameterization based on the twist--2 result by Wandzura
and Wilczek~\cite{Wandzura} that describes
the SLAC data~\cite{E155xpaper} well.

For the error introduced by the uncertainty in the dilution factor, we varied
the cross--normalization between the carbon and ammonia target data by 
an amount of 6\%, consistent with the variations observed for different $W$ and
$Q^2$ ranges and possible differences in the $^{12}$C and $^{15}$N spectra.
This yields an average variation of the dilution factor Eq.~(\ref{dilfac}) of 25\%,
making this error a safe upper bound for {\em all} systematic errors that are
directly proportional to the measured asymmetry.

The CLAS momentum resolution  and reconstruction effects were studied
by moving all data points by 0.02 GeV in $W$ and recalculating the final results.
The effect of this variation on the integrals of $g_1^d$ also gave an upper limit to
systematic errors due to the integration method, which consisted in a simple
sum of all bins
of size $\Delta W = 0.02$ GeV, multiplied by the bin width in $x$.

Other systematic errors were either negligible or have already been described in
the previous section. We note that we do not have a significant {\em systematic}
error from the beam and target polarization product, since they were directly
determined from our data (with minimal theoretical uncertainty).
In particular, the theoretical asymmetry $A_{||}^{elas}$ is only weakly dependent
($\pm 1\%$) on the elastic form factor ratio $G_E/G_M$ for the proton.
 However, the
{\em statistical} error of this method is not negligible and was included
in the total statistical error of the final results.

\section{Results \label{res}}

\begin{figure}[htbp]
\includegraphics[width=8cm]{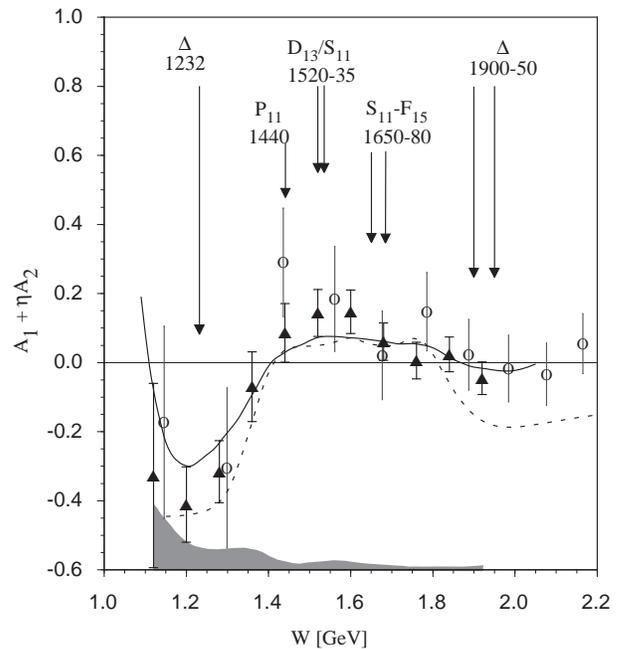}
\caption{
$A_{1}^d+\eta A_{2}^d$ versus $W$ for $Q^{2} = 0.39$  -- $ 0.65$ (GeV/c)$^2$.
Our data points are indicated as triangles with statistical errors only. The size of the systematic
error is indicated by the shaded band at the bottom of the graph. Previous data from 
SLAC E143~\protect\cite{E143Long}
are shown as open
 circles with statistical and systematic errors combined. The positions of
several prominent resonances are indicated by the labeled arrows. The solid line is our
model parametrization of the world data (without nuclear corrections like
Fermi motion and off--shell effects)
and the dashed line is the 
resonant contribution to $A_1^d$ alone (from the code ``AO'').
\label{A1done}}
\end{figure}

\subsection{Virtual photon asymmetries}

We extracted a combination of the virtual photon asymmetries,
 $A_{1}^d + \eta A_{2}^d$, from our data on $A_{||}$ using a
parametrization~\cite{E143R} of the structure function $R$,
%$R ={ \sigma_L \over \sigma_T}$, 
via the relationship
\begin{equation}
A_{||} = D \left( A_{1} + \eta A_{2} \right) ,
\end{equation}
where the virtual photon depolarization factor is given by
 $D = (1- \epsilon E'/E)/(1 + \epsilon R)$ and
$\eta = \epsilon \sqrt{Q^2}/(E- \epsilon E')$ ($\epsilon$ is the virtual photon
polarization parameter, $E$ is the beam energy and $E'$ 
is the scattered electron energy).

\begin{figure}[htbp]
\includegraphics[width=8cm]{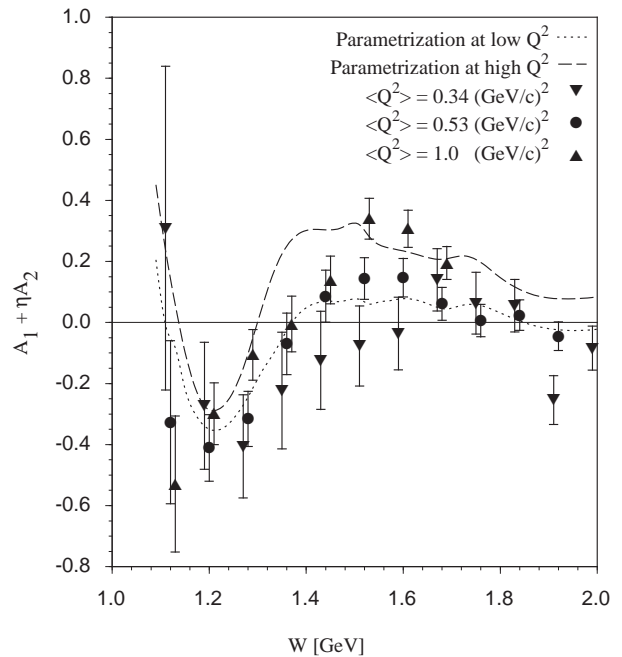}
\caption{
Our data for three different bins in $Q^2$, together with statistical
errors. (Systematic errors are highly correlated between different $Q^2$ bins and
should have only minor effects on the observed $Q^2$-dependence).
The long--dashed line shows our model parametrization
of $A_1^d + \eta A_2^d$ for $Q^2 = 1.0$ (GeV/c)$^2$
and the short--dashed line shows our model for $Q^2 = 0.34$ (GeV/c)$^2$. 
\label{A1dall}}
\end{figure}

The extracted photon asymmetries $(A_{1}^d + \eta A_{2}^d)(W,Q^2)$ 
for three different $Q^2$ bins are listed in
Tables~\ref{ta11}--\ref{ta13}, together with their statistical and full
systematic errors.
We show the results for our intermediate $Q^2$ bin in Fig.~\ref{A1done},
together with previous data from SLAC~\cite{E143Long}
and some model calculations.
A comparison of the three different $Q^2$ bins can be found in
Fig.~\ref{A1dall}.

\begin{table}
\caption{
The measured virtual photon asymmetry $A_1^d + \eta A_2^d$ 
of the deuteron for $Q^2$ = 0.27  --  0.39 (GeV/c)$^2$.
\label{ta11}}
\begin{ruledtabular}
\begin{tabular}{cccc}
 $W$ [GeV] & $A_1^d + \eta A_2^d$ & Stat. Error & Syst. Error \\ 
\hline
1.12 & 0.309 & 0.530 & 0.207 \\
1.20 & -0.273 & 0.208 & 0.061 \\
1.28 & -0.406 & 0.169 & 0.081 \\
1.36 & -0.223 & 0.191 & 0.069 \\
1.44 & -0.124 & 0.161 & 0.028 \\
1.52 & -0.077 & 0.131 & 0.017 \\
1.60 & -0.036 & 0.119 & 0.015 \\
1.68 & 0.140 & 0.102 & 0.023 \\
1.76 & 0.063 & 0.101 & 0.011 \\
1.84 & 0.055 & 0.086 & 0.017 \\
1.92 & -0.254 & 0.080 & 0.028 \\
2.00 & -0.084 & 0.072 & 0.009 \\ 
\end{tabular}
\end{ruledtabular}
\end{table}

\begin{table}
\caption{
The measured virtual photon asymmetry $A_1^d + \eta A_2^d$ 
of the deuteron for $Q^2$ = 0.39 -- 0.65 (GeV/c)$^2$.
\label{ta12}}
\begin{ruledtabular}
\begin{tabular}{cccc}
 $W$ [GeV] & $A_1^d + \eta A_2^d$ & Stat. Error & Syst. Error \\ 
\hline
1.12 & -0.327 & 0.267 & 0.191 \\
1.20 & -0.411 & 0.109 & 0.081 \\
1.28 & -0.316 & 0.090 & 0.061 \\
1.36 & -0.070 & 0.101 & 0.062 \\
1.44 & 0.086 & 0.085 & 0.022 \\
1.52 & 0.144 & 0.068 & 0.025 \\
1.60 & 0.147 & 0.063 & 0.024 \\
1.68 & 0.061 & 0.054 & 0.015 \\
1.76 & 0.006 & 0.053 & 0.011 \\
1.84 & 0.024 & 0.050 & 0.013 \\
1.92 & -0.045 & 0.047 & 0.013 \\
\end{tabular}
\end{ruledtabular}
\end{table}

\begin{table}
\caption{
The measured virtual photon asymmetry $A_1^d + \eta A_2^d$ 
of the deuteron for $Q^2$ = 0.65 -- 1.3 (GeV/c)$^2$.
\label{ta13}}
\begin{ruledtabular}
\begin{tabular}{cccc}
 $W$ [GeV] & $A_1^d + \eta A_2^d$ & Stat. Error & Syst. Error \\ 
\hline
1.12 & -0.529 & 0.223 & 0.125 \\
1.20 & -0.299 & 0.101 & 0.038 \\
1.28 & -0.106 & 0.083 & 0.025 \\
1.36 & -0.005 & 0.091 & 0.046 \\
1.44 & 0.139 & 0.078 & 0.017 \\
1.52 & 0.340 & 0.067 & 0.035 \\
1.60 & 0.307 & 0.061 & 0.038 \\
1.68 & 0.195 & 0.054 & 0.027 \\
1.76 & 0.184 & 0.056 & 0.033 \\
\end{tabular}
\end{ruledtabular}
\end{table}

 Since we didn't measure the asymmetry with the
 target polarization perpendicular to the electron beam ($A_{\perp}$),
we cannot directly extract the asymmetry $A_{1}^d$ or $A_{2}^d$. 
The interference term $A_{2}$ is limited by  
$|A_{2}| < \sqrt{R(A_1+1)/2}$,
where the value of $R$ is around 0.1 - 0.3 at  $Q^{2} = 0.5$ (GeV/c)$^2$~\cite{E143R}
and the typical size of $\eta$ for our experiment
ranges from 0.1 at $W = 2$ GeV to 1.2 right at pion threshold ($W = 1.08$ GeV).
Correspondingly, the asymmetry $A_2^d$
could contribute as much as 0.07 (high $W$) to 0.15 (at threshold) to the asymmetries
shown in Figs.~\ref{A1done} and~\ref{A1dall}. However, according to our parametrization,
this contribution should be more typically 
of order 0.02.

With this caveat, one can conclude that the data shown in Fig.~\ref{A1done}
exhibit the expected behavior for the asymmetry $A_1^d$. In the region of the
$\Delta$(1232) resonance, the asymmetry is strongly negative, and fully compatible
with the expectation $A_1^d = - 0.5$ for the resonance contribution alone. 
Beyond $W=1.4$ GeV, the asymmetry becomes positive,
indicating that helicity--1/2 transition amplitudes begin to dominate even at this rather
low $Q^2$. However, even in the region of the S$_{11}$ resonance the asymmetry
is markedly smaller (around 0.15) than for the proton (around 0.5, see Ref.~\cite{E143Long}),
indicating that for the neutron alone the helicity--3/2 amplitude may still be larger.
Figure~\ref{A1done} also shows the predicted full asymmetry
from our parametrization and a prediction for the resonance contributions to $A_1^d$
alone. The latter is based on the code
``AO''~\cite{BurkertLi}, which uses a fit of exclusive pion electro- and
photoproduction data to parametrize resonant and Born pion production
amplitudes. Apparently, the contribution from the resonances alone
already describes the data well in the region of low to intermediate $W$, while
non--resonant contributions (and maybe a sizable asymmetry $A_2^d$)
are needed at high $W$.
In general, our data agree fairly well with model predictions and the existing
SLAC data. However, they have significantly smaller statistical errors and better
resolution in $W$, as well as coverage down to lower $Q^2$ than the SLAC data.

A comparison of our results for different $Q^2$ (see Fig.~\ref{A1dall}) 
shows a general
trend toward more positive asymmetries for higher $Q^2$, especially in the region of
the S$_{11}$ and D$_{11}$ resonances. This is in agreement with the expected
transition from helicity--3/2 dominance at low $Q^2$ (and especially at the photon
point, where it yields the negative value for the GDH sum rule), and 
helicity--1/2 
dominance at higher $Q^2$. In the limit of very large $Q^2$, the asymmetry
$A_1^d$ in the resonance region should become close to 1, as predicted by pQCD
as well as hyperfine-improved quark models and duality arguments.
A similar behavior is observed for the proton
asymmetries ~\cite{E143Long}.

\subsection{Spin structure function $g_1^d$}

The spin structure function $g_{1}^d(W,Q^2)$ was calculated from the photon
asymmetry $(A_1^d + \eta A_2^d)(W,Q^2)$ for each bin using 
\begin{equation}
\begin{array}{l}
g_{1}^d(W,Q^2) = {{\tau} \over {1+\tau}} \left( A_{1}^d + {1 \over
\sqrt{\tau}} A_{2}^d \right) F_{1}^d(W,Q^2) \\
= {{\tau} \over {1+\tau}} \left( \left( A_{1}^d +\eta A_{2}^d \right)
+ \left(  {1 \over \sqrt{\tau}} - \eta \right) A_{2}^d \right) F_{1}^d(W,Q^2) .
\end{array}
\end{equation}
Here, $F_1^d \approx (F_1^p + F_1^n)/2$
represents the unpolarized structure function
of the deuteron (per nucleon) and $\tau = \nu^2/Q^2$.
Because of the partial cancellation of the two
terms in $\left(  {1 \over \sqrt{\tau}} - \eta \right)$, $g_1^d$ is less sensitive to 
the asymmetry $A_2$. We list our results for $g_1^d$ 
with their statistical and full systematic errors
(including the uncertainty due to $A_2$) in Tables \ref{tg11} - \ref{tg13}.

\begin{table}
\caption{
The spin structure function $g_1^d$ 
of the deuteron for $Q^2$ = 0.27 -- 0.39 (GeV/c)$^2$.
\label{tg11}}
\begin{ruledtabular}
\begin{tabular}{cccc}
$W$ [GeV] & $g_1^d$ & Stat. Error & Syst. Error \\ \hline
1.12	&	0.033	&	0.058	&	0.042	\\
1.20	&	-0.115	&	0.080	&	0.029	\\
1.28	&	-0.172	&	0.074	&	0.033	\\
1.36	&	-0.080	&	0.067	&	0.022	\\
1.44	&	-0.059	&	0.067	&	0.015	\\
1.52	&	-0.040	&	0.078	&	0.010	\\
1.60	&	-0.022	&	0.073	&	0.013	\\
1.68	&	0.112	&	0.074	&	0.020	\\
1.76	&	0.058	&	0.075	&	0.011	\\
1.84	&	0.048	&	0.065	&	0.015	\\
1.92	&	-0.202	&	0.066	&	0.028	\\
2.00	&	-0.070	&	0.066	&	0.012	\\
\end{tabular}
\end{ruledtabular}
\end{table}

\begin{table}
\caption{
The spin structure function $g_1^d$ 
of the deuteron for $Q^2$ = 0.39 -- 0.65 (GeV/c)$^2$.
\label{tg12}}
\begin{ruledtabular}
\begin{tabular}{cccc}
$W$ [GeV] & $g_1^d$ & Stat. Error & Syst. Error \\ \hline

1.12	&	-0.025	&	0.018	&	0.004	\\
1.20	&	-0.088	&	0.025	&	0.017	\\
1.28	&	-0.083	&	0.024	&	0.011	\\
1.36	&	-0.008	&	0.023	&	0.014	\\
1.44	&	0.024	&	0.024	&	0.007	\\
1.52	&	0.075	&	0.029	&	0.012	\\
1.60	&	0.080	&	0.028	&	0.014	\\
1.68	&	0.046	&	0.030	&	0.012	\\
1.76	&	0.016	&	0.031	&	0.011	\\
1.84	&	0.028	&	0.030	&	0.013	\\
1.92	&	-0.016	&	0.031	&	0.013	\\
\end{tabular}
\end{ruledtabular}
\end{table}

\begin{table}
\caption{
The spin structure function $g_1^d$ 
of the deuteron for $Q^2$ = 0.65 -- 1.3 (GeV/c)$^2$.
\label{tg13}}
\begin{ruledtabular}
\begin{tabular}{cccc}
$W$ [GeV] & $g_1^d$ & Stat. Error & Syst. Error \\ \hline

1.12	&	-0.022	&	0.008	&	0.004	\\
1.20	&	-0.029	&	0.011	&	0.003	\\
1.28	&	-0.010	&	0.011	&	0.004	\\
1.36	&	0.003	&	0.011	&	0.007	\\
1.44	&	0.024	&	0.013	&	0.005	\\
1.52	&	0.086	&	0.016	&	0.011	\\
1.60	&	0.089	&	0.016	&	0.013	\\
1.68	&	0.072	&	0.018	&	0.013	\\
1.76	&	0.082	&	0.021	&	0.017	\\
\end{tabular}
\end{ruledtabular}
\end{table}

\begin{figure}[htpb]
\includegraphics[width=8cm]{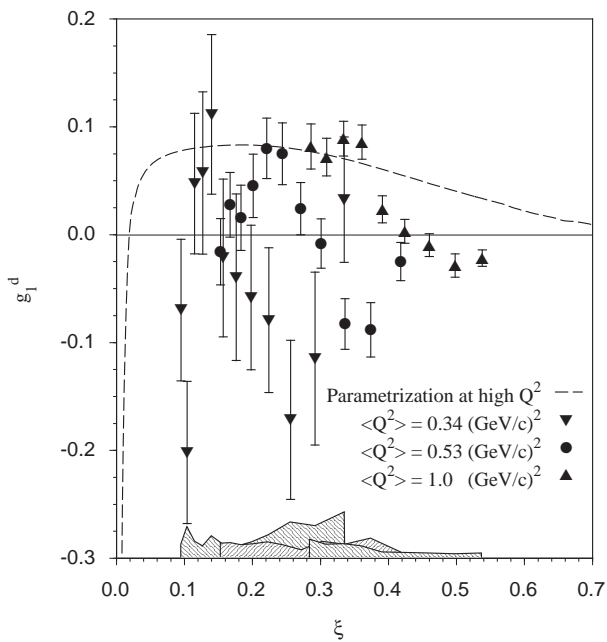}
\caption{
The spin structure function $g_1^d$ for the deuteron
at 3 different values of $Q^2$, plotted against the Nachtmann
variable $\xi$ together with an extrapolation of a fit to the deep inelastic
data at $Q^2$ = 5 (GeV/c)$^2$.
Following standard conventions, all values are
normalized to the number of nucleons in deuterium.
The error bars are statistical only, while the shaded bands
indicate systematic error bars for the three data sets.
\label{g1}}
\end{figure}

In Fig.~\ref{g1}, we show our
results for all three values of $Q^2$,
plotted against the Nachtmann scaling variable $\xi = Q^2 / M(\nu+q)$. This variable
corresponds to Bjorken $x$ at high $Q^2$ while it takes target nucleon mass corrections
into account and therefore reduces ``{\em kinematical} higher twist'' scaling-violating
effects at lower $Q^2$. Together with our data, we also show as reference the prediction
for $g_1^d(\xi, Q^2 = 5$ (GeV/c)$^2)$ from our model. The assumption of local quark-hadron
duality predicts that structure functions like $F_1$ and $g_1$ should, on average, 
approach a universal scaling curve if plotted versus the variable $\xi$, even in the
resonance region. This is confirmed down to rather low $Q^2$ in the case of the
unpolarized structure function $F_2^p$~\cite{NiculescuDual1, NiculescuDual2}.
Apparently, local duality does not work as
well for the {\em polarized} structure function
$g_1^d$ at high values of $\xi$
 where the asymmetry is dominated by the $\Delta$ resonance
and therefore is negative. Overall, the approach to the ``asymptotic value'' for
$Q^2 = 5$ (GeV/c)$^2$ seems to be relatively slow; only our highest $Q^2$ bin
shows fairly good agreement beyond the region of the $\Delta$ resonance.

\subsection{Integrals}

\begin{table}
\caption{The first moments of the spin structure function $g_1^d$
of the deuteron. Following standard convention, the integral is normalized
to the number of nucleons in deuterium.
$Q^2$ is in (GeV/c)$^2$ and $W_{max}$ in GeV.
\label{tgam1}}
\begin{ruledtabular}
\begin{tabular}{ccccccc}
$Q^2$ & $W_{\rm max}$ & Meas. $\Gamma_1$ & Stat. Err. & Sys. Err. 
& Full $\Gamma_1$ & Sys. Err. \\ \hline
%[(GeV/c)$^2$] & [GeV] & -- & -- & -- & --  & -- \\ \hline
0.34 & 2.00 & -0.027 & 0.012 & 0.005 & -0.034 & 0.008 \\
0.53 & 2.00 & -0.008 & 0.004 & 0.002 & -0.013 & 0.007 \\
0.79 & 1.96 & 0.008 & 0.003 & 0.003 & 0.009 & 0.008 \\
1.10 & 1.80 & 0.007 & 0.003 & 0.002 & 0.016 & 0.009 \\
\end{tabular}
\end{ruledtabular}
\end{table}

We calculated the integrals $\Gamma_1^d(Q^2) = \int g_1^d(x,Q^2) dx$
for our results on $g_1^d(x,Q^2)$ over  the (ordinary) Bjorken variable
$x$ for four different $Q^2$ bins, beginning at 
quasi--free pion production threshold ($W$ = 1.08 GeV) up to the
kinematic limit of our data.
(The first two $Q^2$ bins are the same as shown in Tables \ref{tg11} and \ref{tg12},
while we split the last bin into two halves, from $Q^2$ = 0.65 to 0.92 (GeV/c)$^2$
and from $Q^2$ = 0.92 to 1.3 (GeV/c)$^2$.)
 We expect that these integrals are close to
an incoherent average over the individual nucleons (proton and neutron)
in deuterium, reduced by the D-state correction factor $(1-1.5 P_D)$,
where $P_D \approx 0.05$ is the deuteron D-state probability.
The results are shown in the third column of Table~\ref{tgam1} and the
upper kinematic limits for $W$ are listed in the second column.
These
upper $W$ bounds correspond to lower limits of $x = (0.1, 0.15, 0.21, 0.32)$
for the four $Q^2$ bins, respectively. 

We use our model to estimate the contribution
to the integral below these limits and show the resulting ``full'' integrals and their
systematic errors in the last two columns of Table~\ref{tgam1}. These
systematic errors include a contribution from the uncertainty of this extrapolation
to $x = 0$. To estimate this uncertainty, we studied the variation of the low--$x$
contribution according to different fits to the world data;
also, since there are few high--precision data below $x = 0.03$, we
added a systematic error equal to the value of the integral from $x=0$ up to 0.03. 
%new
Due to the large theoretical uncertainty about the shape of the spin structure
functions at very low $x$ and the absence of high-precision data in this region,
the error on this extrapolation may be even larger than indicated by our
systematic error estimate (see below).
%endnew

\begin{figure}[htpb]
\includegraphics[width=8.5cm]{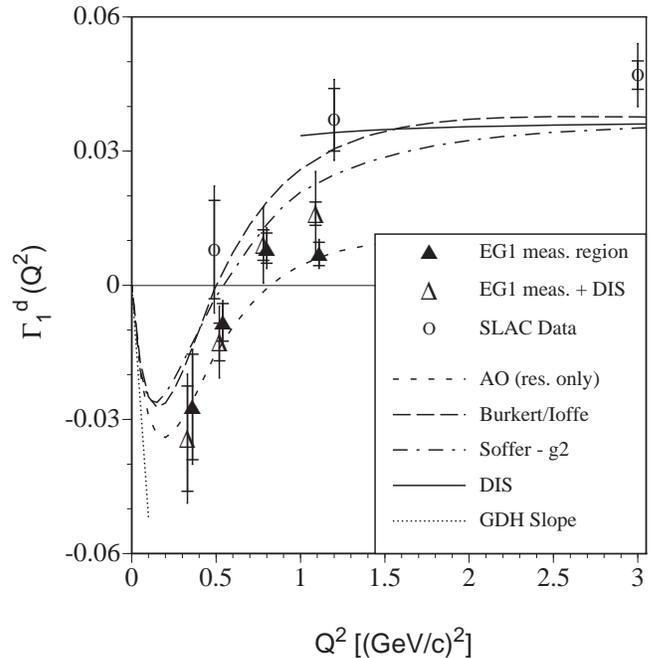}
\caption{
The first moment of the spin structure function $g_1^d$ of the
deuteron (per nucleon). See explanations in text. 
\label{gamma1}}
\end{figure}

Our results for the first moment $\Gamma_1^d(Q^{2})$ of the spin structure 
function $g_1^d$ 
are shown in Fig.~\ref{gamma1}. The solid line at higher $Q^{2}$
is a fit to the world's data in the DIS region including
QCD corrections up to second power in 
the strong coupling constant. The dotted line indicates
the slope for the integral at $Q^2=0$ predicted by the GDH sum rule (we use
the incoherent sum of the results for the proton and for the neutron, normalized
to two).
The short--dashed line is the result from the code ``AO''~\cite{BurkertLi} for 
the contribution from the nucleon resonances only. 
The long--dashed line by Burkert and Ioffe~\cite{BurkertIoffe,BurkertIoffe2}
is the ``AO'' result plus a term that depends smoothly on $Q^{2}$ and 
interpolates between the part that is missing at $Q^{2} = 0 $ to saturate the GDH 
sum rule and the full value of $\Gamma_{1}$ in the high $Q^{2}$ limit.
Fig.~\ref{gamma1}  also shows the prediction from the 
model by Soffer and Teryaev \cite{SofferGam1p,SofferGam1n} (dot--dashed line). 
They use an interpolation of
the integral over the structure function $g_{T} = g_1 + g_2$,
which converges to $\Gamma_{1}$
at high $Q^{2}$ and remains positive down to the photon point
where its slope is given by a combination of the nucleon charge and anomalous
magnetic moment.  They subtract the 
contribution from the integral over $g_{2}$
(which is related to nucleon form factors via the Burkardt--Cottingham sum rule)
 to obtain the integral $\Gamma_{1}$ alone. 
The same authors have recently published a new parametrization of the
proton--neutron difference integral for all $Q^2$~\cite{SofferGam1pn} which
might change the curve for the deuteron shown here.
The solid triangles are based on EG1 data alone
 and the open triangles include the estimated contribution to the integral
from beyond our kinematic limits.
The inner error bars are statistical and the outer error bars
represent the systematic errors added in quadrature.
They include 
the uncertainty on the estimated low-$x$ contribution for the full integrals
(open triangles).

The first conclusion one can draw from Fig.~\ref{gamma1} is that the 
integral over our measured region (essentially the resonance region)
 is in rather good agreement with the
prediction of the ``AO'' parametrization for resonance contributions only.
The data follow the predicted trend from negative values at small $Q^2$, where
the $\Delta$ resonance contributes most of the integral and most other resonances
are also dominated by the helicity--3/2 transition amplitude, to positive
values at higher $Q^2$, where the helicity--1/2 amplitude begins to take over
and the importance of the $\Delta$ is diminished. Since we did not include Born terms
or other non-resonant terms in the curve labeled ``AO'', one can conclude that
these terms must contribute relatively little to the integral over the resonance
region in the case of the deuteron. This may be due to a partial cancellation between
the asymmetry of the proton (which is likely positive for these terms) and 
that of the neutron.

Extrapolating the integral down to $x=0$ seems to change the results only
moderately (in the negative direction at low $Q^2$ and towards more positive
values at higher $Q^2$). This can be understood again as a cancellation between
a strongly negative-going trend of the structure function $g_1^n(x)$ as $x$ goes
to zero and a more positive trend for $g_1^p(x)$, according to existing DIS data and 
NLO analyses~\cite{E154NLO,E155d}.  
However, at present our understanding of the 
behavior of spin structure functions
at very low $x$ is still incomplete, making
this extrapolation rather uncertain (as it is in the
DIS region).
Therefore, the error bars on our open triangles may still underestimate
 that uncertainty. 
The emergence of new information on the low--$x$ behavior of spin structure functions over
 the past five years is responsible for most of the apparent disagreement between our 
quoted results and those from the E143 experiment at SLAC. The integrals over the
 resonance region alone agree fairly well with the SLAC data (to within 1.1 standard
deviations); however, the extrapolation
 beyond $W = 2$ GeV is much more negative for the parametrization used in the present 
analysis and would move the SLAC data points down by about 0.008 and 0.015 at
 $Q^2 = 0.5$ and 1.1 (GeV/c)$^2$, respectively.
%The emergence of new information on the low--$x$ behavior of
%spin structure functions over the past five years
% is partially responsible for the mild discrepancy between
%our quoted results and those from the E143 experiment at SLAC -- if we replace
%the published values for the full integrals with present-day extrapolations for the
%unmeasured region, all SLAC data points would move down by about 0.003 to 0.006.
With this proviso, our data are (marginally) consistent with the SLAC data, but  have
much improved statistical errors and cover lower $Q^2$.

Our data lie somewhat below both phenomenological predictions for the full integral
shown in Fig.~\ref{gamma1}, suggesting 
a slower transition from the negative values near the photon point to the positive
asymptotic value at high $Q^2$. The zero--crossing appears to occur somewhere
between $Q^2 = 0.5$ (GeV/c)$^2$ and $Q^2 = 0.8$ (GeV/c)$^2$, significantly later than
in the case of the proton~\cite{E143Long}.
However, the systematic errors are highly correlated point--to--point so that
the deviation from the predictions by Burkert and Ioffe~\cite{BurkertIoffe,BurkertIoffe2}
and by Soffer and Teryaev \cite{SofferGam1p,SofferGam1n} is not
highly significant.

\section{Summary and Outlook \label{sum}}

In this paper, we report first results on deuterium
for inclusive spin structure functions
 in the nucleon resonance region
from the EG1 program at Jefferson Lab.
These data significantly expand the kinematic coverage and statistical
precision beyond the only previous data from SLAC~\cite{E143Res}.
We find generally reasonable agreement between these two data sets and
various model predictions and parametrizations. In particular, the
importance and the
negative asymmetry of the $\Delta$ resonance is confirmed, as is the
general trend to more positive asymmetries at higher $Q^2$ and $W$.

The spin structure function $g_1^d$ is less positive than for the 
proton case~\cite{E143Long},
 indicating that the neutron contribution is mostly negative in our
kinematic region. While $g_1^d(\xi, Q^2)$ seems to be approaching
the DIS scaling curve for large $W$ and $Q^2$, there are significant
deviations from ``local duality'', again mostly due to the $\Delta$
resonance.

The integral over $g_1^d$ follows the expected trend in general,
rising towards the DIS limit at the highest measured $Q^2$ while
dropping rapidly below 0 towards our lowest $Q^2$ point.
Clearly, neither the kinematic reach (in $W$ and $Q^2$) nor the statistical 
precision of the present data set allow a definite statement about the
validity of (or the approach towards) the GDH sum rule limit.
However, our data constrain
the general trend required of any theory that aims to describe the spin
structure of the nucleon over the full range of length scales, from the
real photon point to the scaling limit.

Spin structure function data on the deuteron,
together with the corresponding proton results,
should in principle allow us to separate the different isospin contributions
to the resonant and non-resonant asymmetries. However, the first run of
EG1 analyzed here did not yield enough statistical precision to make a direct
separation of proton and neutron contributions to the deuteron
asymmetry feasible. However, we plan to submit results on the 
integral $\Gamma_1$ for the neutron
and the proton--neutron difference, extracted from our data on the proton
and the deuteron, in a separate
paper.
In the meantime, the complete EG1 data set
has been collected in a second run, 
which will yield a nearly tenfold improvement in statistics for the deuteron
and a wider coverage towards both lower and higher $Q^2$ and higher $W$.
Once analyzed,
this vastly larger data set will allow us to investigate in detail
resonance electro--production on the neutron and the approach 
of the first moment of $g_1^d$  and $g_1^n$ towards the
GDH sum rule at the real photon point. 

\begin{acknowledgments}
We would like to acknowledge the outstanding effort of the Accelerator, Target Group, 
and Physics Division staff at TJNAF that made this experiment possible.

This work was supported by the U.S. Department of Energy, 
the Italian Istituto Nazionale di Fisica Nucleare,  the U.S. National 
Science Foundation, the French
 Commissariat \`a l'Energie Atomique,
the French Centre National de la Recherche Scientifique,
 and the Korea Science and Engineering Foundation. 
The Southeastern Universities Research Association (SURA) operates the 
Thomas Jefferson National Accelerator Facility for the United States
Department of Energy under contract DE-AC05-84ER40150.

\end{acknowledgments}

%\bibliography{SSF2}

\end{document}